\documentclass[9pt,twocolumn,twoside]{osajnl}
\journal{ol} 
\setboolean{shortarticle}{true} 
\ifthenelse{\boolean{shortarticle}}{\colorlet{color2}{color2b}}{\colorlet{color2}{color2}} 

\title{Line-scan Compressive Raman imaging with spatio-spectral encoding}
\author[1]{Camille Scotté}
\author[1]{Siddharth Sivankutty}
\author[2,3,4]{Randy A. Bartels}
\author[1,5*]{Hervé Rigneault}

\affil[1]{Aix Marseille Univ, CNRS, Centrale Marseille, Institut Fresnel, Marseille, France}
\affil[2]{W.M. Keck Laboratory for Raman Imaging of Cell-to-Cell Communications, Colorado State University, Fort Collins, Colorado 80523, USA
}
\affil[3]{Department of Electrical and Computer Engineering, Colorado State University, Fort Collins, Colorado 80523, USA}
\affil[4]{School of Biomedical Engineering, Colorado State University, Fort Collins, Colorado 80523, USA}
\affil[5]{Lightcore Technologies, Marseille, France}
\affil[*]{Corresponding author: herve.rigneault@fresnel.fr}
\dates{Compiled \today}

\begin{abstract}
We report a line-scanning imaging modality of Compressive Raman technology with a single-pixel detector. The spatial information along the illumination line is encoded onto one axis of a digital micromirror device, while spectral coding masks are applied along the orthogonal direction. We demonstrate imaging and classification of three different chemical species.
\end{abstract}
\begin{document}
\maketitle

\begin{figure}[h!]
\includegraphics[width=\linewidth]{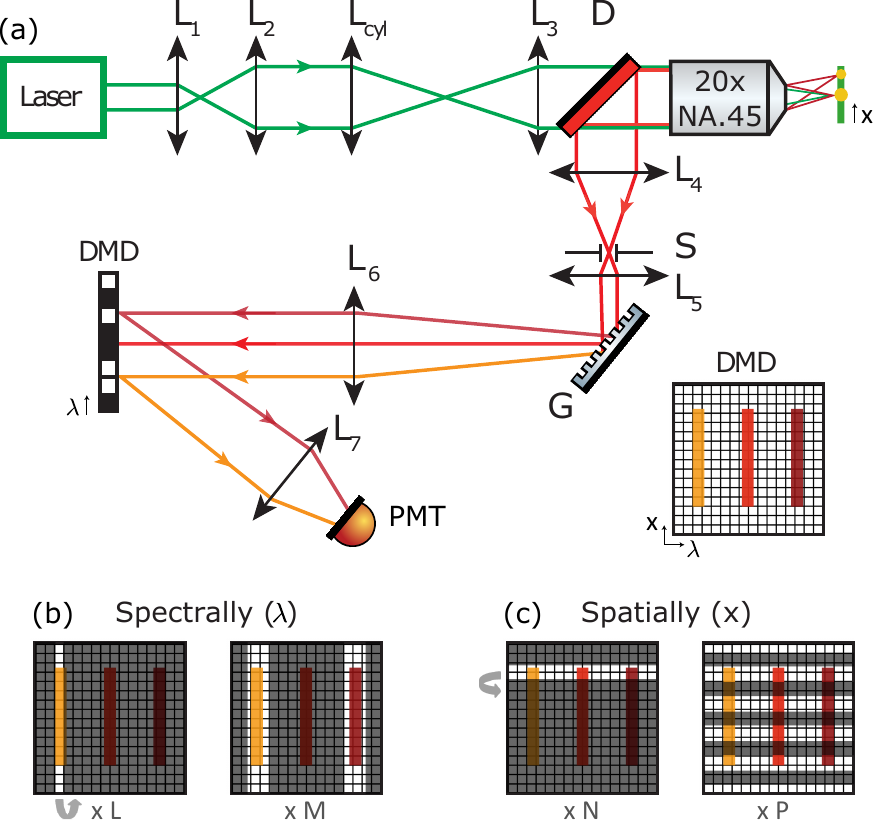}
\caption{(a) Schematic of the experimental setup.  $L_1 - L_6$ -  convex lenses with focal lengths $50$~mm, $150$~mm, $150$~mm, $150$~mm, $100$~mm, $150$~mm and $50$~mm, respectively. $L_7$ -  combination of 2 lenses that image the DMD into the PMT with $\times$ 3 de-magnification. $L_{cyl}$ - cylindrical lens with focal length $150$~mm, D - dichroic mirror, S - confocal slit, G - amplitude grating (600~lines/mm), PMT - photomultiplier tube. (b) The spectral axis gives access to the Raman spectrum, and allows to project optimized spectral filters. (c) The spatial axis gives access to the spatial information along the line focus.}
\label{fig:Expt}
\end{figure}

Spontaneous Raman spectroscopy permits characterization of molecular species with high specificity in a label-free manner. 
Typically, the Raman inelastically scattered light is spectrally dispersed and collected onto an array detector - for several positions of the sample - resulting in a Raman hyperspectral image. This measurement of a complete vibrational Raman spectrum per spatial pixel, coupled with the weak Raman scattering cross-section and detector array noise, requires lengthy acquisition times and generates large data sets. 
In situations where hyperspectral measurements simply aim to map the spatial distribution of known molecules, such an acquisition process is highly inefficient. Instead of unmixing the spectral data in a post-processing step to detect molecular species and/or estimate their concentrations \cite{Keshava2002, Palkki2010}, higher acquisition efficiencies can be achieved by encompassing compressive techniques in the acquisition process. In Compressive Raman Technology (CRT), the measurement is directly designed to estimate quantities of interest (e.g., molecular concentrations), rather than deducing them from complete hyperspectral
measurements \cite{Wilcox2012, Wilcox2013, Buzzard2013, Refregier2018, Scotte2018, Cebeci2018, Sturm2019}.
This is achieved by replacing the array detector by a single-pixel detector combined with a fast programmable optical filter, typically a digital micromirror device (DMD). Based on the \textit{a priori} knownledge of the Raman spectra of the pure molecular species contained in the sample, these filters select accurately chosen spectral components and combine them into the detector (Fig. \ref{fig:Expt}). Such CRT developments have led to spontaneous Raman imaging with pixel dwell times down to 30 $\mu$s~\cite{Wilcox2012, Rehrauer2017}, which is orders of magnitude faster than state-of-the-art Raman hyperspectral imaging \cite{Scotte2018}.\\ 
To further improve CRT, we recently demonstrated a CRT line-scan strategy based on spatial frequency-modulated illumination imaging \cite{Futia2011}, referred to as CRiSPY \cite{Scotte:19}.
As opposed to recent spatial encoding approaches \cite{Galvis-Carreno2014, Thompson2017, Choi2017}, CRiSPY does not use an spatially-resolved detector but combines both spatial and spectral information into a single-pixel detector. It encodes the 1D spatial information into temporal frequencies, by modulating the illumination line with chirped cosines imprinted on a rotating disk \cite{Futia2011, Scotte:19}. In the present Letter, we explore an alternative line-scan CRT strategy that does not require an external modulator, but instead makes use of the bi-dimensionality of the DMD (Fig.~\ref{fig:Expt}): (i) The Raman spectral information is dispersed across the $\lambda-$axis, and the DMD encodes this information with dedicated spectral filters; (ii) The spatial information along the line-focus is encoded onto the DMD x$-$axis, with Hadamard matrices shifted to contain only positive values. 
We demonstrate line-scan CRT with proof-of-concept experiments for chemically specific classification of molecular species, and give evidence that this approach is beneficial over point-scanning CRT in a number of cases.





\begin{figure}[H]
\includegraphics[width=\linewidth]{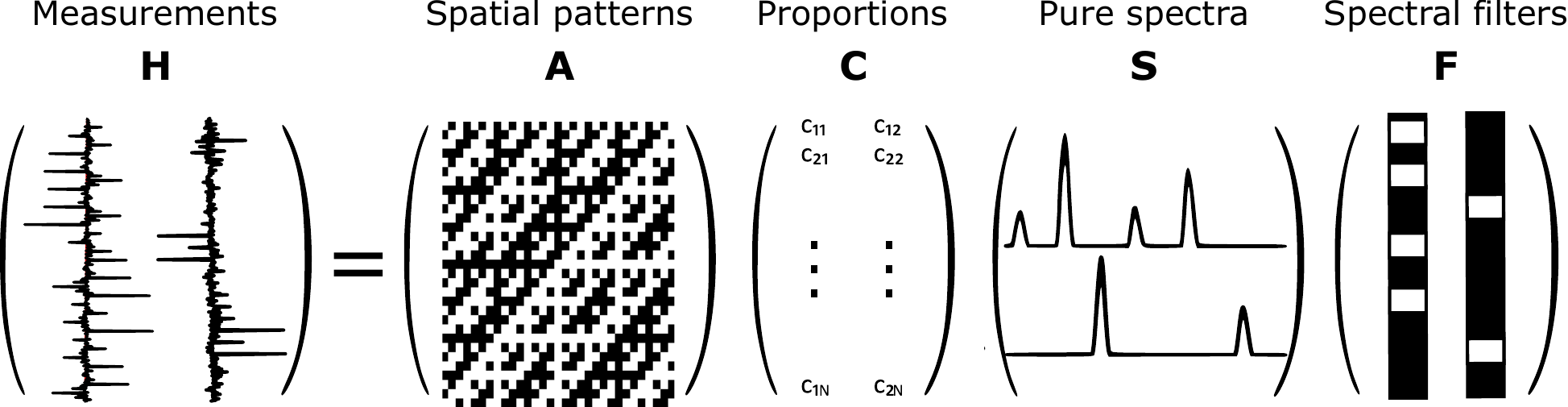}
\caption{Visual representation of Eq. \ref{eqn:globalmatrix1}, with $Q = M = 2$.}
\label{fig:Formalism}
\end{figure}

\noindent In line-scan CRT, both spatial (along the line-focus) and spectral information is encoded in the signal measured onto the single-pixel detector. The aim is to estimate the relative abundances (proportions) of a set of $Q$ molecular species (with known Raman spectra) in each of the $N$ resolved points along the illumination line (Fig.~\ref{fig:Formalism}). We denote $P$ as the number of DMD patterns along the x$-$axis ($p = 1...P$), $Q$ as the number of pure chemical species present in the sample ($q = 1...Q$), $M$ as the number of spectral filters ($m = 1...M$) and $L$ as the number of resolved energy bins along a Raman spectrum ($l = 1...L$).
The $N\times Q$ matrix $\mathbf{C}$ is the quantity to estimate, i.e., the spatial distribution of the pure chemical species proportions.
Each element of $\mathbf{C}$, $c_q(x_n)$, specifies the proportions of the $q^{th}$ pure chemical species contained in the resolved point $x_n$ of the illumination line. 
The  $P\times N$ matrix $\mathbf{A}$ contains the $P$ patterns displayed along the DMD x$-$axis.
The $L\times M$ matrix $\mathbf{F}$ contains the $M$ spectral filters $\mathbf{f_m}$.
The $Q\times L$ matrix $\mathbf{S}$ contains the known Raman spectra of the pure chemical species. 
The measurements matrix, $\mathbf{H}$, contains the number of counts measured when displaying binary masks onto the DMD. Each mask is formed by the Kronecker product of a row of $\mathbf{A}$ and a column of $\mathbf{F}$ (Fig.~\ref{fig:Expt}). Each column of H is recorded by keeping the spectral filter fixed, and scanning through the rows of $\mathbf{A}$ to obtain a complete set of spatial projections. 
Assuming the generated signal is linear to the excitation intensity and ignoring constant terms, the measurement process can be expressed as: 
\begin{align}
\mathbf{H} = \mathbf{A} \mathbf{C} \mathbf{S} \mathbf{F} = \mathbf{A} \mathbf{C} \mathbf{G}^T
\label{eqn:globalmatrix1}
\end{align}
where $\mathbf{G^T}= \mathbf{S} \mathbf{F}$. 

Although the spatial and spectral dimensions could be considered conjointly, we handle them distinctly in this Letter. In the spectral domain, the filters $\mathbf{f}_m$ are calculated from the pure Raman spectra $\mathbf{s}_q$ with the same optimisation procedure as in \cite{Refregier2018}.
In the spatial domain, we choose the matrix $\mathbf{A}$ to be a modified Hadamard matrix with positive coefficients (S-matrix). The S-matrix of size $N$ is obtained by removing the first row and column of a Hadamard matrix of size $N+1$, and changing the ones to zeros and minus ones to ones \cite{Harwit1979}. It is square ($P = N$), invertible, and its binary nature complies with the DMD design and its use at fast frame rates.
We perform the estimation in two steps. First, we estimate a spatial line image of Raman intensities, for each spectral filter. In other words, we seek to estimate the global matrix $\mathbf{C} \mathbf{G}^T $, which elements $\eta_m(x_n)$ represent the Raman intensity in each pixel $x_n$, for the spectral filter $\mathbf{f}_m$. Since our measurements are shot-noise limited \cite{Scotte2018} and that $\eta_m(x_n) \geq 0 $, we perform this estimation using the classical EM algorithm (Richardson-Lucy), which seeks to maximize the Poisson likelihood under positivity constraints \cite{Gurioli2014, Lucy1974, Shepp1982}. For relatively sparse objects, its gives better performances in terms of mean-square error as compared to least-square estimation (data not shown), as previously reported in the literature \cite{Bialkowski1998,Fuhrmann2004, Harmany2012b}. We will report more precisely on this effect in a forthcoming publication. The resulting estimate is denoted $\hat{\eta}_m(x_n)$. 
After a transpose operation, the estimation of the species proportions reduces to a 1D CRT problem \cite{Refregier2018}:
\begin{align}
\hat{\boldsymbol\eta}^T(x_n) &= \mathbf{G} \mathbf{c}^T(x_n) 
\label{eqn:CRT2}
\end{align}
with $\hat{\boldsymbol\eta}(x_n)=(\hat{\eta}_1(x_n),...,\hat{\eta}_M(x_n))$ and $\mathbf{c}(x_n)=(c_1(x_n),...,c_Q(x_n))$.
If $\mathbf{G}^T\mathbf{G}$ is not singular, the proportions are finally estimated via least-square estimation \cite{Refregier2018}:
\begin{align}
\mathbf{\hat{c}}^T(x_n) &= \big[ \mathbf{G}^\intercal \mathbf{G} \big]^{-1} \mathbf{G}^\intercal \hat{\boldsymbol\eta}^T(x_n)\
\label{eqn:CRTestimation}
\end{align}



A simplified  schematic of the experimental setup is depicted in Fig.~\ref{fig:Expt}. On the illumination side, a continuous wave laser operating at 532~\textit{nm} (Verdi, Coherent Inc) is spectrally filtered and expanded. The beam is brought to a line focus onto the sample plane with a combination of cylindrical lens, plano-convex lens and microscope objective (Nikon 20x, 0.5 NA). 
On the detection side, the scattered light from the object is relayed onto a confocal slit. A combination of dichroic mirror and notch filter ensures only the Raman signal is retained. Next, it is dispersed with a blazed grating ($600~mm^{-1}$, Thorlabs), and the
spatially dispersed wavelength components are imaged onto the DMD (V-7001, Vialux -$1024\times768$ mirrors). The DMD $\lambda-$axis, in conjunction with the grating, acts as a programmable spectral filter.  Since the line focus is imaged onto the DMD ($\approx$ 22.5 de-magnification), its x$-$axis offers control over the corresponding spatial dimension of the object [Inset Fig.~\ref{fig:Expt}]. When the DMD pixels are in the 'ON' state, the signal impinging on these pixels is deflected into a photon-counting PMT (H7421-40, Hamamatsu), while the rest is sent into a beam dump.
A piezoelectric stage scanner (P517, Physik Instrumente GmBH) holding the sample is used to scan the y$-$ axis, yielding 2D images. 


\noindent In this configuration, the spatial resolution along x$-$ is limited by the imaging system and the size of one DMD mirror. In the experiments, the DMD mirrors are binned 2-by-2 along x$-$, resulting in an equivalent spatial resolution of about 1.2 $\mu$m. The line length on the illumination plane, together with cropping from diverse optical elements, limits the field-of-view along x$-$ to about 220 $\mu$m. On the DMD plane, the spatial extent of the line image only represents $1/3$ of the entire DMD size, but the S-patterns are displayed over the full DMD x$-$range ($P = 511$). In the y$-$ direction, the spatial resolution is about 1.5 $\mu$m (extent of the line focus on the sample).  The spectral resolution is estimated to $40~cm^{-1}$, limited by the grating and the imaging system. This allows to bin the DMD mirrors 8-by-8 along $\lambda -$. 
Spatial calibration was performed by scanning a bead of known size and matching its profile onto the DMD. The effective FOV was estimated by recording the Raman signal from a spatially homogeneous sample (microscope glass slide) and scanning it along the DMD x$-$ axis.  \\

\begin{figure}[h!]
\includegraphics[width=\linewidth]{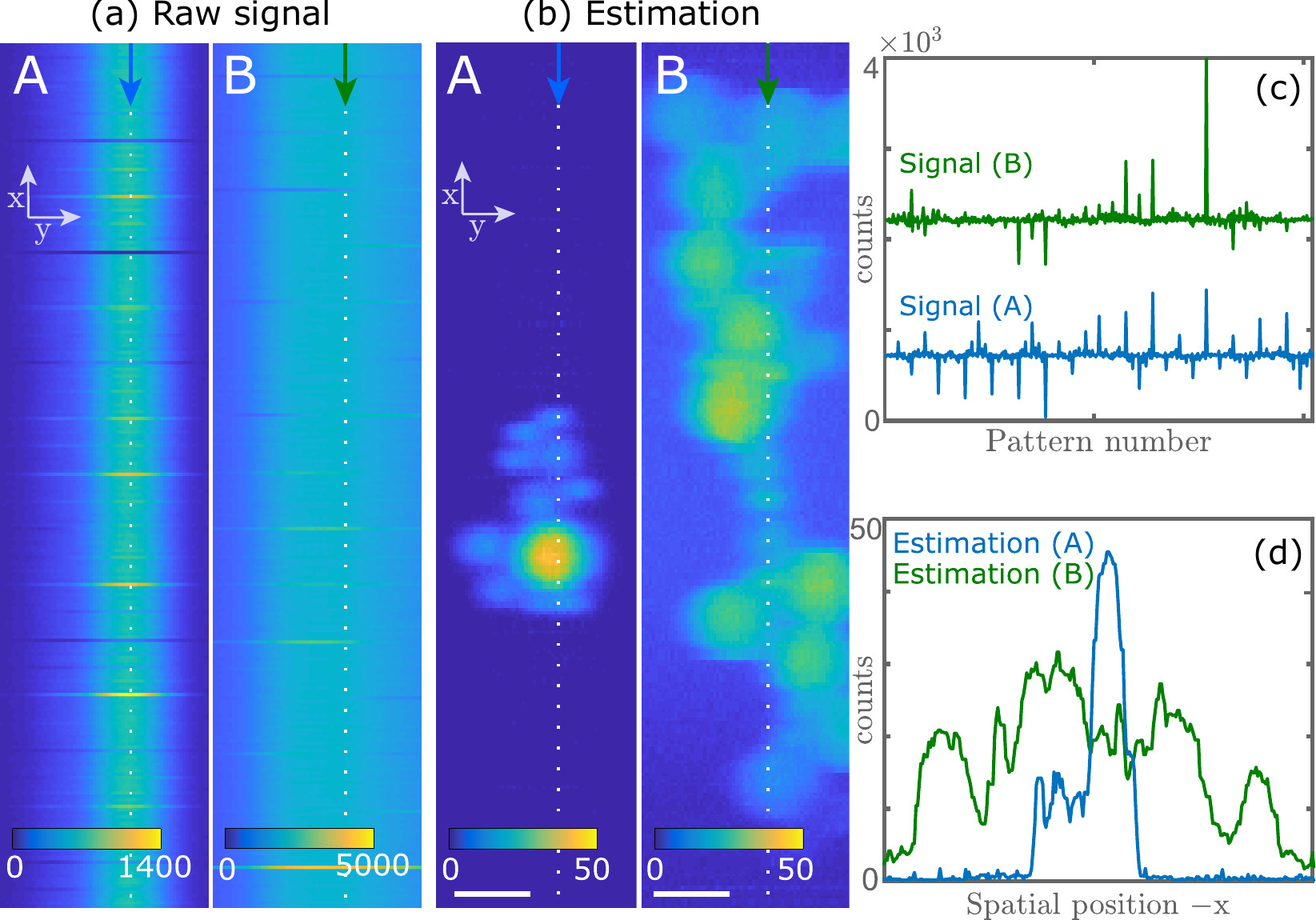}
\caption{(a) Signal after projection of S-patterns along each line x$-$, for sample A and B. (b) Estimated images obtained with the EM algorithm. (c-d) Cross-sections along the dotted lines of (a, b). Results are averaged over 50 measures, and cropped to show only the FOV. Scale bar = 30$\mu$m. x$-$: Line-scan axis, y$-$: Piezo-scan axis. Colorbars units: counts.}
\label{fig:Data_demowhitelight}
\end{figure}

The first experiment consisted in acquiring Raman line images with no spectral selectivity ($f_{lm} = 1 \forall l,m $) of two mixtures of beads (Sigma Aldrich). $30~\mu$m polystyrene beads (PS), $20~\mu$m polymethylmethacrylate (PMMA) and  $12~\mu$m melamin resin (MR) were dispersed in two different manners Fig.~\ref{fig:Data_demowhitelight}(b): (A) in a sparse way, onto a CaF$_2$ coverslip; (B) in a dense way, onto a glass coverslip. The total laser power on the illuminated line was set to 3.6 mW (1.1 $\times$ 10$^{-5}$ W$/ \mu$m$^2$), so that the total count rate lies in the linearity regime of the detector (1.5 $\times$ 10$^6$ Hz). Fig. \ref{fig:Data_demowhitelight}(a) shows the signal obtained upon the projection of each S-pattern for 10 ms, averaged over 50 measurements, and Fig. \ref{fig:Data_demowhitelight}(b) the associated estimated Raman line images. The signal along the lines of sample B is higher due to the presence of more chemical species along x$-$ [Fig. \ref{fig:Data_demowhitelight}(a, c)]. No further post-processing was performed, except from the removal of some systematic periodic artefacts on the estimated images, arising from yet unindentified technical issues.

\begin{figure}[h!]
\includegraphics[width=\linewidth]{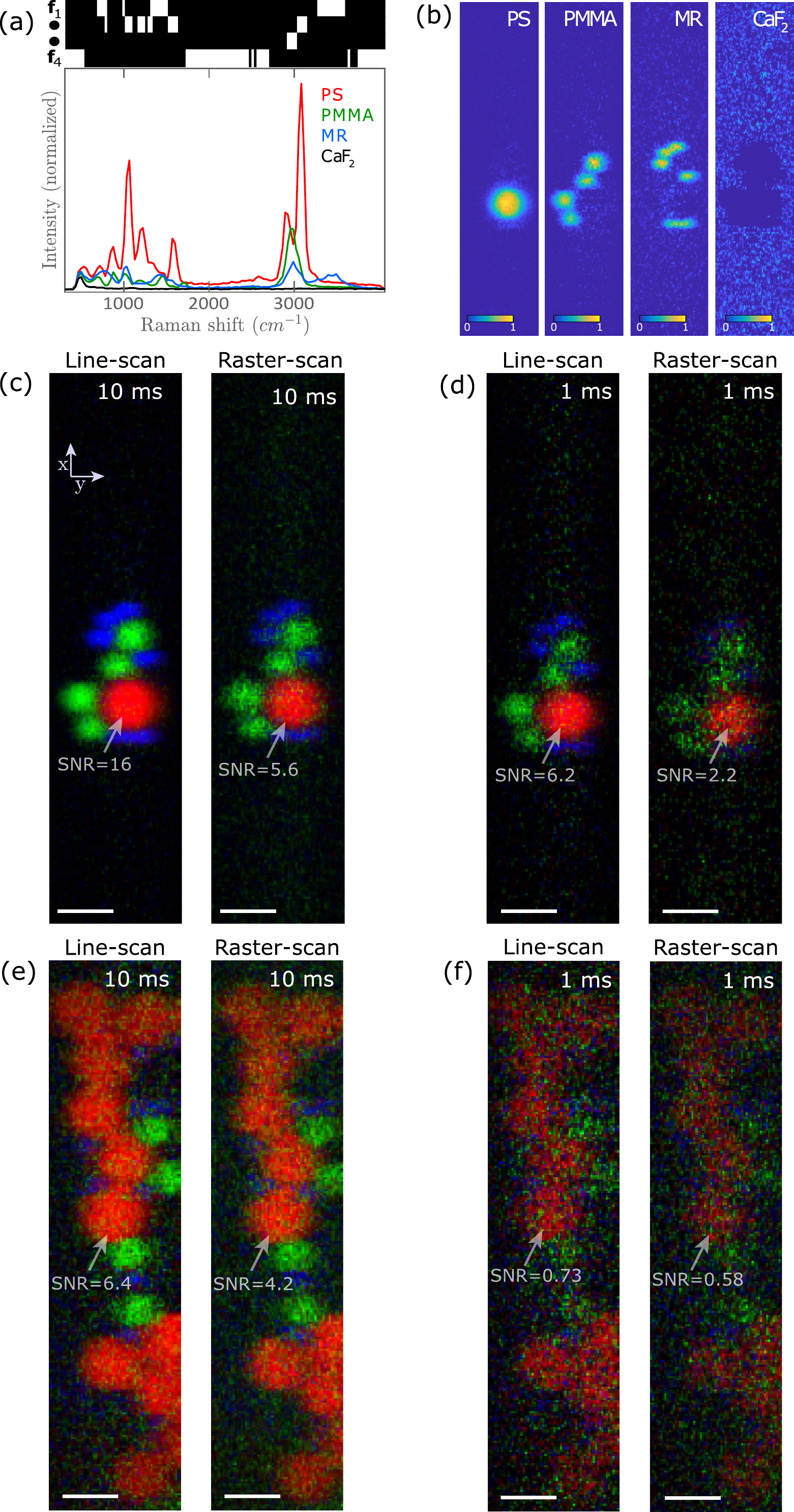}
\caption{(a) Reference spectra of the 4 chemical species, and associated spectral filters, for sample A. (b) Estimated proportion maps with line-scan. (c-f) Visualization of proportion maps of PS (red), PMMA (green) and MR (blue), for line-scanning and raster-scanning.}
\label{fig:CRTResults}
\end{figure}

\noindent Next, we performed CRT experiments to obtain Raman images with chemical selectivity. The reference spectra were measured with using the DMD x$-$axis as a virtual pinhole \cite{Scotte:19} [Fig.~\ref{fig:Expt}], and the spectral filters $\mathbf{f}_m$ optimised as in \cite{Refregier2018, Scotte2018}. The resulting spectra and filters for sample A ($Q=M=4$) are shown in Fig.~\ref{fig:CRTResults}(a). 
In CRT experiments, the total laser power along the line was chosen higher (10.6 mW - 3.3 $\times$ 10$^{-5}$ W$/ \mu$m$^2$) than for pure Raman line imaging described above, because of the lower signal due to the spectral filtering. 
The 4 line-images, corresponding to each spectral filter, are estimated, followed by the proportion estimation in each pixel along the line via Eq. \ref{eqn:CRTestimation}. The resulting proportion maps for sample A, thresholded to [0 1], are shown in Fig.~\ref{fig:CRTResults}(b). The composite RGB maps are obtained by combining the normalised proportion maps of the three beads types.
The results are shown for integration times of 10 ms and 1 ms per projected S-pattern, with no averaging. As a comparison, the same experiment was performed with raster-scanning the DMD pixels along x$-$ [Fig~\ref{fig:Expt}(c-left)], with same irradiance and integration times. This is formally equivalent to scanning the sample plane with a point-focus. Visually, the line-scan measurements seem superior to the raster-scan measurements, but this disparity is more striking for the sparse sample A than for sample B. The difference in image quality is quantitatively confirmed by the SNR values on the central pixels of a PS bead (Fig. \ref{fig:CRTResults} (e-g)), defined as $ SNR_{PS} = \langle \hat{c}(x_{PS})\rangle / \sigma_{PS}$, where $\langle \hat{c}(x_{PS})\rangle$ and $\sigma_{PS}$ are the mean and standard deviation of the estimated proportions, respectively. The SNR improvement over raster-scanning is about 3 for object A, and 1.5 for object B.\\ 

\begin{figure}[h!]
\includegraphics[width=\linewidth]{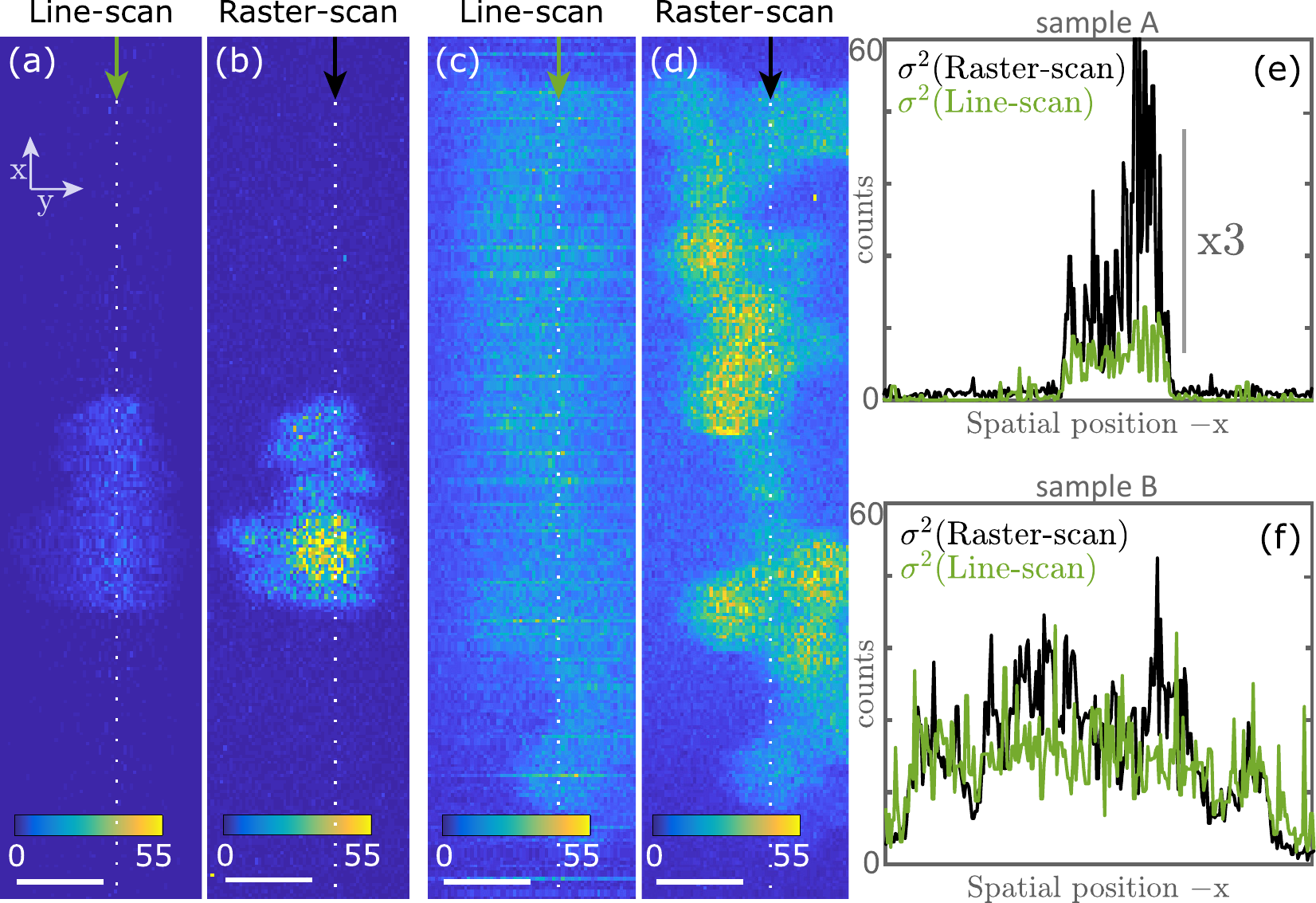}
\caption{Variance images (50 measurements), for line-scanning and raster-scanning. (a-b): sample A. (c-d): sample B. (e-f) Cross-sections along the dotted lines of (a-d).}
\label{fig:Variance}
\end{figure}

To better assess the efficiency of the presented line-scanning technique over point-scanning, we compare the estimation variances of the corresponding Raman line images, with no spectral selectivity (Fig. \ref{fig:Variance}). 
For sample A, the variance on the object is lower for line-scanning than for raster-scanning. 
For example, on the PS bead along central line of the sample, the variance $\sigma^2$ is about 3 times lower, leading to a SNR improvement of $\sqrt{3}$ (SNR on pixel $i$: $SNR_i = \langle \hat{\eta}(x_i) \rangle / \sigma_i$) . For sample B, line-scanning leads to a relatively uniform variance along the line which is similar to raster-scanning. This leads to a limited SNR improvement. 
These results are consistent with the CRT results of Fig. \ref{fig:CRTResults}. They show that, in terms of SNR, line-scanning seems mostly beneficial for the sparse sample A, and brings a limited improvement to the denser sample B. This stems from the shot-noise limited nature of our measurements \cite{Scotte2018}. Here, the noise scales with the square-root of the number of photons arising from the illuminated pixels across the line \cite{Studer2012}. In addition, the signal from pixel $x_n$ may be affected by photon noise coming from the average Raman signal along the entire line. This results in an object-dependent SNR improvement, which is expected to increase for sparser objects \cite{Harmany2012b, Bialkowski1998,Fuhrmann2004, Studer2012}. This is in contrast to measurements where detectors exhibit signal-independent additive noise: In this case, the same S-multiplexing strategy would lead to an SNR improvement of $\sqrt{N}/2$ \cite{Harwit1979}, i.e. of 10 times, for both objects. These aspects will be investigated in more details in a forthcoming publication. \\
\noindent From this short analysis, we surmise that our line-scan CRT approach would bring a SNR advantage over point-scanning CRT, and thus be faster, when samples are relatively sparse 
or when the experimental configuration exhibits signal-independent noise contributions. We emphasize that the spectral filtering induced by CRT renders the samples more sparse, and thus further improves the SNR, as long as the chemical species spectral overlap is limited. 
The SNR and speed gain will then depend on the spatial and spectral structure of the sample along the multiplexed line. \\

In this Letter, we have demonstrated a simple approach for line-scan CRT with S-multiplexing, with proof-of-concept experiments. While a line-scan strategy includes some drawbacks such as non-uniform illumination profile \cite{Hutchings2009} or potential loss in spatial or spectral resolution, we surmise line-scan CRT to be beneficial over point-scanning CRT for SNR in many circumstances. While the mentioned speeds do not compete with point-scanning CRT, they are expected to be faster with approaches such as compressive-sensing \cite{Candes2006b, Berto2017}. In addition, the results could further be improved by filters design and optimisation taking into account the spatial and spectral domain conjointly.


\paragraph{Funding Information} We acknowledge financial support from the Centre National de la Recherche Scientifique (CNRS), Institut Carnot STAR (IRAC 2018), Aix-Marseille University A$^\ast$Midex (noANR-11-IDEX-0001-02 and A-M-AAP-ID-17-13-170228-15.22-RIGNEAULT), ANR grants France Bio Imaging (ANR-10-INSB-04-01) and France Life Imaging (ANR-11-INSB-0006) infrastructure networks and Plan cancer INSERM PC201508 and 18CP128-00. C.S. has received funding from the European Union’s Horizon 2020 research and innovation program under the Marie Skłodowska-Curie grant agreement No713750. 
\paragraph{Acknowledgements} The authors ackowledge fruitful discussion with P. Réfrégier and F. Galland.

\bigskip
\bibliography{CRISPY}
\ifthenelse{\equal{\journalref}{ol}}{%
\clearpage
\bibliographyfullrefs{CRISPY}
}{}

\ifthenelse{\equal{\journalref}{aop}}{%
\section*{Author Biographies}
\begingroup
\setlength\intextsep{0pt}
\begin{minipage}[t][6.3cm][t]{1.0\textwidth} 
  \begin{wrapfigure}{L}{0.25\textwidth}
    \includegraphics[width=0.25\textwidth]{john_smith.eps}
  \end{wrapfigure}
  \noindent
  {\bfseries John Smith} received his BSc (Mathematics) in 2000 from The University of Maryland. His research interests include lasers and optics.
\end{minipage}
\begin{minipage}{1.0\textwidth}
  \begin{wrapfigure}{L}{0.25\textwidth}
    \includegraphics[width=0.25\textwidth]{alice_smith.eps}
  \end{wrapfigure}
  \noindent
  {\bfseries Alice Smith} also received her BSc (Mathematics) in 2000 from The University of Maryland. Her research interests also include lasers and optics.
\end{minipage}
\endgroup
}{}
\end{document}